\documentclass{article}
\usepackage{fullpage}
\usepackage{wrapfig}
\usepackage[colorlinks,linkcolor=blue,urlcolor=blue,citecolor=blue]{hyperref}
\usepackage{amsmath,amsfonts}
\usepackage{graphicx,color}
\usepackage{listings}
\usepackage{verbatim}

\newcommand{\dealii}{{\textsc{deal.II}}}

\newcommand{\trilinos}{{\textsc{Trilinos}}}
\newcommand{\aspect}{\textsc{Aspect}}
\newcommand{\petsc}{\textsc{PETSc}}

\begin{document}


\title{Clone and graft:\\ Testing scientific
  applications as they are built}

\author{Bruno Turcksin\footnote{Department of Mathematics, Texas A\&M
    University, College Station, TX 77843-3368, USA, \url{turcksin@math.tamu.edu}}
\and
   Timo Heister\footnote{Mathematical Sciences, O-110 Martin Hall, Clemson University,
     Clemson, SC 29634-0975, USA, \url{heister@clemson.edu}}
\and
   Wolfgang Bangerth\footnote{Corresponding author. Department of Mathematics, Texas A\&M
     University, College Station, TX 77843-3368, USA, \url{bangerth@math.tamu.edu}}}

\maketitle

This article describes our experience developing and maintaining automated tests for scientific applications. The main idea evolves around building on already existing tests by cloning and grafting. The idea is demonstrated on a minimal model problem written in Python.

This article was originally planned as part of a book on testing scientific software\footnote{See \url{https://github.com/swcarpentry/close-enough-for-scientific-work} for details.}. The code and sources for the article are available at \url{https://github.com/tjhei/clone-graft-paper}.

\section{Introduction}

Our group has been building scientific software for more than a decade and a
half by now. Over the years we have developed many procedures that help us
test our software extensively but our expertise was mostly in the development
of \textit{libraries} for numerical methods -- specifically, the \dealii{}
library for finite element computations (see
\url{http://www.dealii.org/} and \cite{BHK07,BK99m}). Libraries are of course the foundation of
almost every single scientific code, starting with ``simple'' ones such as
BLAS and LAPACK to very much more complex ones such as \petsc{}, \trilinos{}, or
\dealii{} \cite{petsc,trilinos,trilinos-web-page}.

Developing testing schemes for libraries is reasonably well understood. So,
when we ventured in the area of building \textit{applications} for scientific purposes
on top of the libraries we had created, we had initially thought that one can
use the same approaches as one uses for libraries -- but this turned out to
not work.

Libraries are relatively easy to test. They export large numbers of
functions; one can write a simple \texttt{main()} function that sets up a
couple of data structures, calls one of these functions individually, and
verifies the correctness of the output; do
this for every function or class in the library and you've got a good
testsuite. Such tests are
often called \textit{unit tests} and are typically complemented by integration
tests to verify that combinations of classes work well together. Over the
years we have written about 3,000 of these that are run in multiple
configurations with every single commit to \dealii{} -- on multiple machines,
with different compilers and dependencies, and several times a day. Despite
all of the experience we have gained building this machinery around \dealii{},
the same strategy does not work for whole applications because applications do
not usually export many possible entry points. Rather, they usually only
present a single interface -- the command line, a graphical user interface, or
input files -- and then run through a large fraction of the entire code base
in computing output. One cannot typically just call a single function in an
application by presenting a magic input file, in the same way as one would
call a function in a library by presenting a magic \texttt{main()} function in
a unit test. There are of course more advanced ways of testing features of an
application in isolation by replacing other parts of the code during testing by so-called
mock or stub objects. We will not discuss this here though.

Consequently, different approaches are necessary. In the following, we will
lay out what we have learned from building and testing the \aspect{} code for the
geodynamics community over the past few years. In the following, let us first
talk briefly about what \aspect{} does and how the way it is used is
prototypical for scientific codes (Section~\ref{sec:aspect}). At the same time,
\aspect{} is a very complex code and just installing it is non-trivial;
consequently, it does not serve the purpose of this book well and we will
present a simple model problem and model code in
Sections~\ref{sec:model-problem} and \ref{sec:first-steps} that we
will then use throughout the rest of the chapter. Sections~\ref{sec:testing}
and \ref{sec:extending}
discuss how we write tests for whole applications, and how this process
happens as we continue to develop the code -- using the
\textit{clone and graft} technique of creating test
cases. Section~\ref{sec:practice} then deals with how real practice interferes
with the ideas discussed in previous sections, and we conclude and summarize
in Section~\ref{sec:conclusions}.

\section{\aspect{}: the Advanced Solver for Problems in Earth's ConvecTion}
\label{sec:aspect}

\begin{wrapfigure}{R}{0.44\textwidth}
  \begin{center}
    \vspace*{-24pt}
    \includegraphics[width=0.42\textwidth,height=0.42\textwidth]{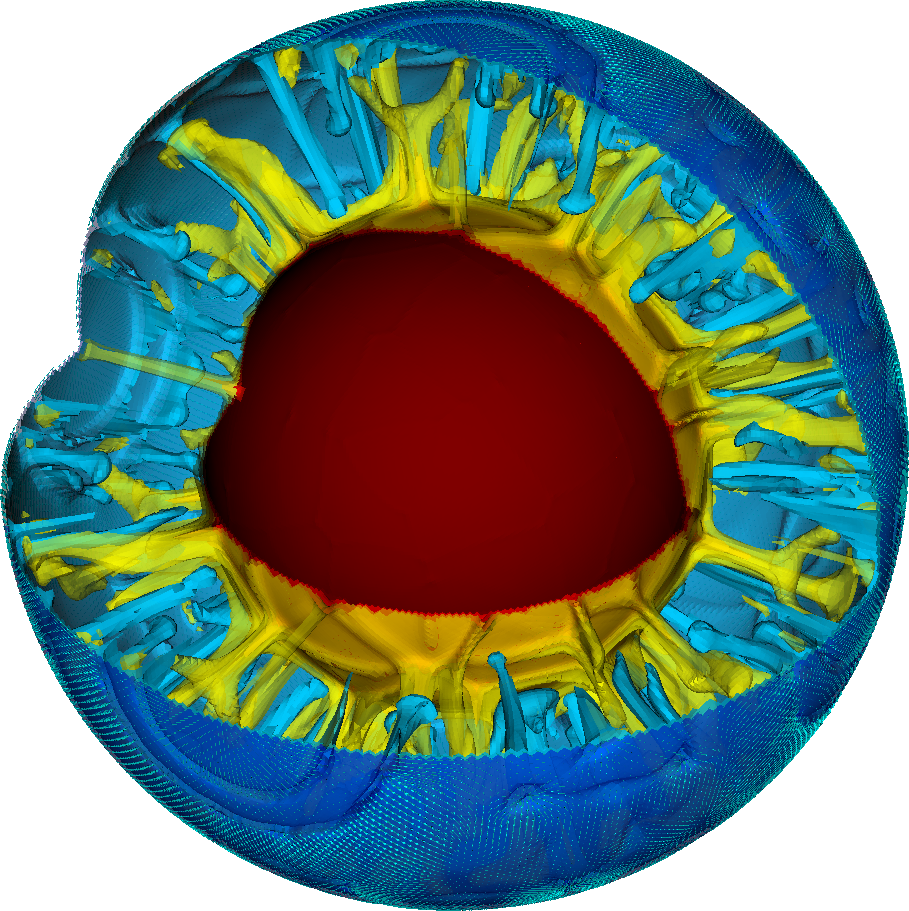}
    \vspace*{-12pt}
  \end{center}
  \caption{\it Snapshot of the temperature field of an \aspect{} simulation
    showing convection in the Earth mantle.}
  \vspace*{-3mm}
  \label{fig:aspect}
\end{wrapfigure}
The code that made us think about how to test whole applications is called
\aspect{} -- short for the Advanced Solver for Problems in Earth's ConvecTion
\cite{KHB12,aspectmanual,aspectweb}. It is a program that simulates how material moves around the Earth
mantle (i.e., the region between the metallic core of Earth and the plates at
the surface on which we live). While the material in the mantle is solid rock,
it is hot and under enormous pressure and can deform with velocities of a few
centimeters per year. On time scales of millions of years, it therefore
behaves like a fluid. Since it is heated from below and cooled from above, it
shows the same kind of behavior as a pot on the stove: blobs of hot material
rise up, cool at the surface, and blobs of cool material fall down. In other
words, it convects. The details of trying to simulate this on a computer go
beyond what we want to discuss here, but it is worth showing a picture and
linking to the videos at \url{http://www.youtube.com/embed/j63MkEc0RRw} and
\url{http://www.youtube.com/embed/EJJ6f4hmDPU}, if only because they are pretty.

\aspect{} is a large code: At the time of writing this chapter, it has 297
source files with 69,704 lines of code. What's more, it builds on the
\dealii{} library that has some 500,000 lines of code, the \trilinos{} library
with 3,500,000 lines, and a few other but smaller libraries. At the same time,
this size is not uncommon for large scientific codes, and certainly not for
commercial codes.

Like many academic codes, \aspect{} has no graphical user interface but
instead is driven by input files; it is also designed in such a way that it is
easy to add functionality through plugins, i.e., self-contained source files
that simply implement a class derived from some base class and that is then
registered in \aspect{}'s plugin registry at run time. Because this is the way
users interact with this code, it is also the framework within which we have
to approach testing: tests need to consist of specially crafted input files
and/or plugins. Plugins go beyond what we can discuss here, but we will show
you that it is actually quite simple to write tests using input files.

\section{Make it simple for me: A model problem}
\label{sec:model-problem}

We're not going to try demonstrating our approach using \aspect{} -- that
would be far to complex to install and deal with. Rather, we're going to show
how it works with a much simpler code that just deals with a model problem but
that we will write with the same approach towards input handling and testing
as \aspect{}.

So this is what we're going to consider: Imagine you are dealing with two
baseballs that
are connected by a spring. If you throw them, what are their trajectories?
Newton's law says that for each of the two balls, mass times acceleration
equals the force on the body. Let's assume for a moment that there is no air
friction and that the spring is massless, then the ordinary differential
equation (ODE) that describes the motion of each of the two bodies ($i=1,2$) is
\begin{align}
  \label{eq:ode}
  m_i 
  \underbrace{\mathbf x_i''(t)}_{\text{acceleration}}
  &=
  \underbrace{m_i \mathbf g}_{\text{gravity}}
  -
  \underbrace{D (\|\mathbf x_2(t) - \mathbf x_1(t)\| - L)}_{\text{magnitude of
    spring force}}
\underbrace{\mathbf d_i}_{\text{direction of spring force}},
\\
\label{eq:ic1}
  \mathbf x_i(0) &= \mathbf x_{i,0},
\\
\label{eq:ic2}
  \mathbf x_i'(0) &= \mathbf v_{i,0},
\end{align}
where $\mathbf x_i(t)$ is the position of the $i$th body at time $t$, $\mathbf
g=(0,0,-g)^T$ is the gravity acceleration, $D$ is the spring constant of
the spring that connects the two balls, and $L$ is the rest length of the
spring. The direction vector $\mathbf d_i$ is
$\mathbf d_1 = \frac{\mathbf x_2(t) - \mathbf x_1(t)}{\|\mathbf x_2(t) - \mathbf
    x_1(t)\|}$
and
$\mathbf d_2 = \frac{\mathbf x_1(t) - \mathbf x_2(t)}{\|\mathbf x_1(t) - \mathbf
    x_2(t)\|}=-\mathbf d_1$ for the two masses.
The second and third equations are the necessary initial conditions
for this second order differential equation and denote the initial position
$\mathbf x_{i,0}$ and initial velocity $\mathbf v_{i,0}$.

\section{A first program}
\label{sec:first-steps}

Let's implement a program that can solve this problem. Because we want it
to be simple and accessible, we're going to use Python as it's widely used and
because it comes with a number of tools that are going to make our life
simpler by keeping the program short.%
\footnote{In order to run this program, you will have to have the Python
  NumPy, SciPy, and matplotlib packages installed on your system, along with
  the basic Python interpreter.}
Note that testing is even more important in dynamically typed languages like
Python, because you do not have the compiler to help you find bugs at compile 
time.
  
The program we'll show you in a second reads its input parameters from a file in JSON format%
\footnote{While certainly more sane compared to XML, JSON is not the prettiest format to write input files in and we
  don't recommend you use it for your programs (it is typically intended for
  exchange of data between programs, not between humans and computers). But it
  will serve for our purposes here primarily because Python has a built-in
  parser for it that makes our program pleasantly free of the tedium of
  parsing input files.} (see
\url{http://en.wikipedia.org/wiki/JSON}).
Here is an example that you can find in the file \texttt{tests/testcase-1.json}:
\begin{lstlisting}[frame=single,numbers=left,basicstyle=\footnotesize]
{
"initial position"   : [ [1,2,3], [4,5,6] ],
"initial velocity"   : [ [0,0,0], [1,1,1] ],
"masses"             : [ 13.5, 29.75 ],
"spring constant"    : 42,
"spring rest length" : 2.25
}
\end{lstlisting}
In JSON, the name of a parameter is to the left of the colon and its value is
to the right; name/value pairs are separated by commas. The value of a
parameter can be a plain number, or it can be an array if enclosed in
brackets. Arrays can also be nested, which is what we use here for the initial
positions and velocities of our two bodies, each of which lives in
three-dimensional space. With this explanation, it is clear that the input file
specifies properties of the two masses as $\mathbf x_{1,0}=(1,2,3)^T, \mathbf
x_{2,0}=(4,5,6)^T, \mathbf v_{1,0}=(0,0,0)^T, \mathbf v_{2,0}=(1,1,1)^T,
m_1=13.5, m_2=29.75$ and $D=42, L=2.25$ for the spring that connects them, 
thereby completely determining everything in equations
\eqref{eq:ode}--\eqref{eq:ic2}.

To implement a program that can deal with this input, we first need to
describe how Python solves ODEs. Basically, it requires us to reformulate it
as a first order system $\mathbf y'(t)=\mathbf f(t,\mathbf y), \mathbf
y(0)=\mathbf y_0$, and tell the integrator about $\mathbf f$ and $\mathbf
y_0$. We do this by introducing variables for the velocities $\mathbf v_i=\mathbf x'_i$ and then
setting $\mathbf
y=(x_{1,1},x_{1,2},\ldots,x_{2,3},v_{1,1},v_{1,2},\ldots,v_{2,3})^T$. $\mathbf
f$ then consists of the $\mathbf v_i$ and the right hand sides of
\eqref{eq:ode} divided by the $m_i$. With this, the code looks like this:

\begin{lstlisting}[frame=single,basicstyle=\footnotesize,numbers=left,language=Python]
import sys
import json
import numpy
import scipy.integrate

# Read the name of the input file from the command line, and read options from
# the file:
assert len(sys.argv) == 2, 'Please provide an input file.'
with open(sys.argv[1], 'r') as f:
    settings = json.loads(f.read())

# Then retrieve the various parameters from what we just read:
D = settings['spring constant']
L = settings['spring rest length']
m = settings['masses']
[x1_0, x2_0] = settings['initial position']
[v1_0, v2_0] = settings['initial velocity']

# describe the differential equation as a first order ODE:
y0 = [x1_0[0], x1_0[1], x1_0[2], x2_0[0], x2_0[1], x2_0[2],
      v1_0[0], v1_0[1], v1_0[2], v2_0[0], v2_0[1], v2_0[2]]

def f(t, y):
    p1 = y[0:3]
    p2 = y[3:6]
    v1 = y[6:9]
    v2 = y[9:12]

    g = [0., 0., -9.81]

    dist = numpy.linalg.norm(p2-p1)
    a1 = g - D*(dist-L) * (p1-p2)/dist/m[0]
    a2 = g - D*(dist-L) * (p2-p1)/dist/m[1]
    return numpy.concatenate([v1, v2, a1, a2])


# Next create an object that can integrate the ODE numerically:
start_time = 0.
end_time   = 5
integrator = scipy.integrate.ode(f)
integrator.set_integrator('vode', rtol=1e-6)
integrator.set_initial_value(y0, start_time)

# With this, do the integration step by step, appending values to an array in
# each step:
t_values = [start_time]
y_values = numpy.array([y0])
while integrator.successful() and integrator.t < end_time:
    integrator.integrate(end_time, step=True)
    t_values.append(integrator.t)
    y_values = numpy.vstack((y_values, integrator.y))

# Having done so, output the number of time steps and the final positions:
print "time steps:", len(t_values)
print "final position:", y_values[-1,0:3], y_values[-1,3:6]
\end{lstlisting}

\begin{wrapfigure}{R}{0.3\textwidth}
  \begin{center}
    \vspace*{-24pt}
    \includegraphics[width=0.28\textwidth]{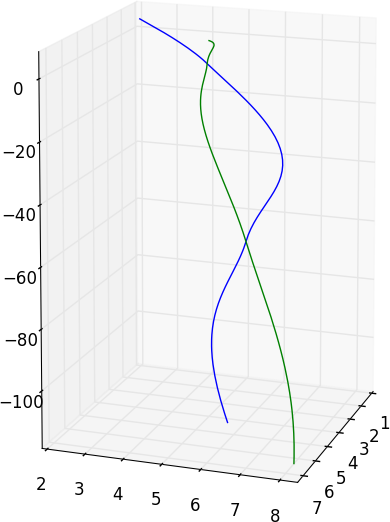}
    \vspace*{-12pt}
  \end{center}
  \caption{\it Trajectories for the two balls with the input file from
    Section~\ref{sec:first-steps}. Note that axes are not shown at the same
    scale.}
  \vspace*{-15mm}
  \label{fig:testcase-1}
\end{wrapfigure}

When you run the program on this data, it will compute the trajectory of the
two connected balls over a number of seconds and print their final positions
to the console.
Of course, having a plot of their trajectories would be nice too, and you can
get that by adding the following lines to the bottom of the script:
\begin{lstlisting}[frame=single,basicstyle=\footnotesize,numbers=left,language=Python]
import matplotlib.pyplot
from mpl_toolkits.mplot3d import Axes3D
fig = matplotlib.pyplot.figure()
canvas = fig.gca(projection='3d')
canvas.plot(y_values[:,0], y_values[:,1], y_values[:,2],
            label='body 1')
canvas.plot(y_values[:,3], y_values[:,4], y_values[:,5], 
            label='body 2')
canvas.legend()
matplotlib.pyplot.show()
\end{lstlisting}
The output this creates is shown in Fig.~\ref{fig:testcase-1}.

\section{Testing for perpetuity}
\label{sec:testing}

So how does one test a code like this? There are really two answers to
this. The first has to do with ensuring that the functionality we have just
implemented is correct. The second is how we can ensure that the functionality
we have implemented today will still work correctly tomorrow after
implementing more features.

We often have an intuitive idea of how to do the former. For example, we could
note that the center of mass at the end time $T=5$ should have fallen a
distance of $\delta z=\frac 12 g T^2=\frac{9.81 \cdot 25}{2}\textrm m\approx
123\textrm{m}$, well in agreement with what Fig.~\ref{fig:testcase-1} shows. We
could also compare with another code. Or we could look at special cases for
which we have 
an analytic solution. For our model problem, one could for example prescribe
initial velocities so that the bodies are at rest and initial positions so
that the two bodies are separated by the rest length of the spring; the two
bodies then simply fall straight down and we can easily compute on a piece of
paper where they should be at time $t$. The following input file
(\texttt{tests/testcase-2.json}) does this:
\begin{lstlisting}[frame=single,numbers=left,basicstyle=\footnotesize]
{
"initial position"   : [ [0,0,0], [1,0,0] ],
"initial velocity"   : [ [0,0,0], [0,0,0] ],
"masses"             : [ 1, 2 ],
"spring constant"    : 1,
"spring rest length" : 1
}
\end{lstlisting}

In the end, verifying whether the program is correct is problem dependent. Let
us not go too deep into this because the second answer -- how to ensure that
functionality continues to work -- is actually much more important in practice
even though few people realize this when they start developing software. The
issue essentially boils down to this: We need mechanisms that ensure that
the functionality we have verified works correctly today, does not break when
we implement new functionality tomorrow. This is necessary because good programmers
understand that they will make mistakes; what makes them good programmers is
that they develop ways to minimize the impact by helping them find problems
quickly. 

There is a lot of research on bugs in programs. Fundamentally, the essence of
this research is that it cannot be avoided altogether when programming. On the
other hand, the best programs are those that make it easy to spot bugs and fix
them quickly. The worst programs are those that are not tested; an empirically
true statement is that programs that are not tested produce wrong answers --
where the emphasis is on the fact that they \textit{do} produce wrong answers,
not that they \textit{may produce} wrong answers. Experienced programmers all
attest to this statement. Only bad or inexperienced programmers will assume
that they do not introduce new bugs in old parts of their programs.%
\footnote{In fact, what seems to set good programmers apart more than many
  other factor is not that they create fewer bugs -- which they do, though not
  dramatically so -- but that they understand that they create bugs and that
  they change their behavior to accommodate for it, for example by developing
  a habit of rigorous testing that helps them find these bugs quickly.}

The importance of testing relies in the cost of fixing problems. If we keep
implementing new functionality, eventually some old feature will stop
working. If we have no tests that detect this quickly, this may go unnoticed
for a while but eventually we will try to re-run an old computation with the
new version of the code and it will not work any more. So where to start? What
have we changed in the intervening months that may cause this? There is likely
a lot of new code we have written since then, and we may have forgotten about
a fair share of the details since then. It will take a long time to figure out
where exactly the problem is. On the other hand, if we had a testsuite that we
run after essentially every change, we will be notified immediately that something stopped
working -- with a relatively small number of changes made since
the test last worked, and everything still fresh on our mind, finding the bug
should be a much quicker issue.

So how do we achieve this? The general idea is that we save the test cases we
have previously used to convince ourselves that a new part of the program was
correct. To make this work, we have to have three things:
\begin{enumerate}
\item The input file for a previously verified testcase.
\item The output when running the code on this input file; having verified the
  feature, we know or at least believe that this output is correct.
\item A way to run a new version of the code on the output file, to compare
  its current output against the stored output, and to raise a flag if they
  are not the same.
\end{enumerate}
The first two are really not very complicated -- create a directory
\texttt{tests/} in your project and create two files in it, say
\texttt{testcase-1.json} and the corresponding output
\texttt{testcase-1.reference}. When it comes time to test at a later time, step 3
above can be as easy as running commands such as
\begin{lstlisting}[frame=single,basicstyle=\footnotesize,numbers=left,language=csh]
  python spring.py testcase-1.json > testcase-1.output
  diff testcase-1.reference testcase-1.output
\end{lstlisting}
You would repeat all of this for \texttt{testcase-2.json}, of course.
If the \texttt{diff} program produces no output, then the two files are the
same and your program computed the same answer on this input file as it did
back when you created the \texttt{testcase-1.reference} file. If there is a
difference, then you know that you just broke one of the testcases. If you
purposefully changed it -- for example because you changed the output format,
or because you fixed behavior you know was erroneous and have verified that it
is now correct, then copy the new output over the old reference file. If you did not
purposefully change what the program computes, then you know it's time to break out the debugger.

\section{More functionality}
\label{sec:extending}

We write tests because we expect that our code base will continue to change --
to implement new functionality, or to fix existing bugs. So let's do this:
assume we now also want to include air friction on the two balls, and we will
use that the air friction force is proportional to the square of the speed and
in the opposite direction of the velocity. Then our model
\eqref{eq:ode} needs to change to the following:
\begin{align}
  \label{eq:ode2}
  m_i 
  \mathbf x_i''(t)
  &=
  m_i \mathbf g
  -
  D \left(\|\mathbf x_2(t) - \mathbf x_1(t)\| - L\right) \mathbf d_i
  \underbrace{-C_i \; \|\mathbf x_i(t)\| \; \mathbf x_i(t)}_{\text{friction force}}.
\end{align}
The new term contains constants $C_i$ that denote air friction coefficients of
the two balls. These coefficients can be computed from the cross section of
the ball, their surface roughness, the viscosity of the air, and other
factors, but this is not important here -- we will simply read them from the
input file.

If we now require that every input file contains an array of friction
coefficients $C_i$, we can no longer use old input files. A better strategy is
to define default values for all non-required parameters and overwrite them
with the values in the input file if specified. A backward compatible choice
of input parameters for the $C_i$ is to give them a default value of zero
because in that case equation \eqref{eq:ode2} equals the previous version,
\eqref{eq:ode}.

Implementing all of this leads to the following extension of the program shown
above%
\footnote{This is the file \texttt{spring.py} in the directory corresponding to
this paper.}, with the new parts in lines 19--20 and 35--36:
\begin{lstlisting}[frame=single,numbers=left,basicstyle=\footnotesize,language=Python]
import sys
import json
import numpy
import scipy.integrate

# Read the name of the input file from the command line, and read options from
# the file:
assert len(sys.argv) == 2, 'Please provide an input file.'
with open(sys.argv[1], 'r') as f:
    settings = json.loads(f.read())

# Then retrieve the various parameters from what we just read:
D = settings['spring constant']
L = settings['spring rest length']
m = settings['masses']
[x1_0, x2_0] = settings['initial position']
[v1_0, v2_0] = settings['initial velocity']

# the default friction is zero for both objects:
[C1, C2] = settings.get('air friction coefficient', [0, 0])

# describe the differential equation as a first order ODE:
y0 = [x1_0[0], x1_0[1], x1_0[2], x2_0[0], x2_0[1], x2_0[2],
      v1_0[0], v1_0[1], v1_0[2], v2_0[0], v2_0[1], v2_0[2]]

def f(t, y):
    p1 = y[0:3]
    p2 = y[3:6]
    v1 = y[6:9]
    v2 = y[9:12]

    g = [0., 0., -9.81]

    dist = numpy.linalg.norm(p2-p1)
    a1 = g - D*(dist-L) * (p1-p2)/dist/m[0] - C1*numpy.linalg.norm(p1)*p1
    a2 = g - D*(dist-L) * (p2-p1)/dist/m[1] - C2*numpy.linalg.norm(p2)*p2
    return numpy.concatenate([v1, v2, a1, a2])


# Next create an object that can integrate the ODE numerically:
start_time = 0.
end_time   = 5
integrator = scipy.integrate.ode(f)
integrator.set_integrator('vode', rtol=1e-6)
integrator.set_initial_value(y0, start_time)

# With this, do the integration step by step, appending values to an array in
# each step:
t_values = [start_time]
y_values = numpy.array([y0])
while integrator.successful() and integrator.t < end_time:
    integrator.integrate(end_time, step=True)
    t_values.append(integrator.t)
    y_values = numpy.vstack((y_values, integrator.y))

# Having done so, output the number of time steps and the final positions:
print "time steps:", len(t_values)
print "final position:", y_values[-1,0:3], y_values[-1,3:6]

# graphical output:
if False:
    import matplotlib.pyplot
    from mpl_toolkits.mplot3d import Axes3D
    fig = matplotlib.pyplot.figure()
    canvas = fig.gca(projection='3d')
    canvas.plot(y_values[:,0], y_values[:,1], y_values[:,2],
                label='body 1')
    canvas.plot(y_values[:,3], y_values[:,4], y_values[:,5], 
                label='body 2')
    canvas.legend()
    matplotlib.pyplot.show()
\end{lstlisting}

How do we test this new functionality? First, we should check whether the old input files still
work:
\begin{lstlisting}[frame=single,basicstyle=\footnotesize,numbers=left,language=csh]
  python spring.py testcase-1.json > testcase-1.output
  diff testcase-1.reference testcase-1.output
  python spring.py testcase-2.json > testcase-2.output
  diff testcase-2.reference testcase-2.output
\end{lstlisting}
With the code that accompanies this paper, \texttt{diff} reports no difference
between the old and the new output files, so this is reassuring.

Next, we need to write new tests for the new functionality. This is where the
\textit{clone and graft} technique finally comes into play: take one of the
existing tests (i.e., \textit{clone} it) and add to it the minimal number of
parameters to exercise the new functionality (i.e., \textit{graft} onto
it). So let's take the simpler of the two previous testcases,
\texttt{tests/testcase-2.json}, copy it to \texttt{tests/testcase-3.json} and
add to it something about air friction: 
\begin{lstlisting}[frame=single,numbers=left,basicstyle=\footnotesize]
{
"initial position"         : [ [0,0,0], [1,0,0] ],
"initial velocity"         : [ [0,0,0], [0,0,0] ],
"masses"                   : [ 1, 1 ],
"spring constant"          : 1,
"spring rest length"       : 1,
"air friction coefficient" : [ 1, 1 ]
}
\end{lstlisting}
To verify that the results we get are correct, we can use that in this test
case the masses and friction coefficients of the two balls are the same, the
two balls are separated by a distance equal to the rest length of the spring,
and they start at rest. Then they will simply fall as if there was no spring,
and we know that they will approach a terminal speed $v=\sqrt{mg/C}$ where
downward gravity and upward air friction cancel. For the current case, this
would be $v=\sqrt{9.81}$ because we have chosen $m=1, g=9.81, C=1$.

Is one new testcase enough? The answer is rarely yes. For example, it is easy
to make indexing errors and choosing masses and friction coefficients equal
would not allow us to detect errors where, for example, we used
\texttt{friction[1]} in the code where we intended to use \texttt{friction[i]}.
So let's copy \texttt{testcase-3.json} to
\texttt{testcase-4.json}:
\begin{lstlisting}[frame=single,numbers=left,basicstyle=\footnotesize]
{
"initial position"         : [ [0,0,0], [1,0,0] ],
"initial velocity"         : [ [0,0,0], [0,0,0] ],
"masses"                   : [ 1, 2 ],
"spring constant"          : 1,
"spring rest length"       : 1,
"air friction coefficient" : [ 1, 1 ]
}
\end{lstlisting}
Here, the second ball has a greater mass, but the same air friction
coefficient (think of it as denser but with the same diameter). How do we know
whether our code is correct for this case as well? This may be more difficult
to establish, but it isn't always necessary. For example, we could look at the
output and realize that the second ball falls faster and, because it is
connected to the first by a spring, then starts to drag the first behind it as
it falls. Such qualitative verification is sometimes the best one can do. At
the same time, while it may not guarantee that the code is indeed correct, it
will nonetheless serve as a useful test in the future to ensure that existing
functionality does not break as a result of further development -- in the long
term the more important of the two reasons to write tests.

Because it is so easy to create tests, it may be worthwhile cloning
\texttt{testcase-3.json} a second time into
\texttt{testcase-5.json} where now we just set the friction coefficients
differently:
\begin{lstlisting}[frame=single,numbers=left,basicstyle=\footnotesize]
{
"initial position"         : [ [0,0,0], [1,0,0] ],
"initial velocity"         : [ [0,0,0], [0,0,0] ],
"masses"                   : [ 1, 1 ],
"spring constant"          : 1,
"spring rest length"       : 1,
"air friction coefficient" : [ 1, 2 ]
}
\end{lstlisting}
Now the two balls have the same mass but the second one has a higher drag
coefficient (think of the first ball as smooth and the second one as covered
in fur). Our ``smell check'' would now be to verify that the second ball drags
behind the first.

If we're reasonably convinced that the output is correct, then save the output
in files \texttt{tests/testcase-3.reference} through
\texttt{tests/testcase-5.reference} for use in future testing.

\section{Practical considerations}
\label{sec:practice}

\subsection{When clone and graft fails}

\begin{wrapfigure}{R}{0.2\textwidth}
  \begin{center}
    \vspace*{-24pt}
    \includegraphics[width=0.1\textwidth]{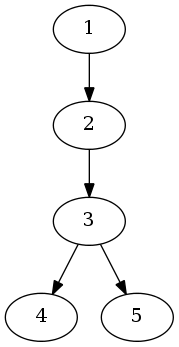}
    \vspace*{-12pt}
  \end{center}
  \caption{\it Inheritance graph for the 5 tests discussed in previous sections.}
  \vspace*{-3mm}
  \label{fig:inheritance}
\end{wrapfigure}
The scheme outlined above represents an inheritance relationship where every
test is based on a previous one. For the 5 testcases discussed above, the
inheritance graph is shown in Fig.~\ref{fig:inheritance}. It is, of course, a
\textit{tree graph}.

But this does not always work. There are cases where you will have to write a
new test from scratch. To give an example of where this was necessary,
consider \aspect{}: Imagine that at the beginning, it could only compute
convection in a box. We would have testcases that described different
temperature boundary conditions on this box, and various other things. But
when implementing a spherical geometry, none of these testcases quite fit:
they all had to say what the temperature is on the six faces of the box, they
made the implicit assumption that gravity points downward instead of inward,
etc. It was time to start with a new testcase. If, at a later time you then
clone and graft from this new testcase, then it becomes the root of a separate
tree (let's call it an \textit{exemplar} testcase) and the inheritance graph
will turn into a \textit{forest graph} (a collection of tree graphs).

\begin{wrapfigure}{L}{0.45\textwidth}
  \begin{center}
    \vspace*{-24pt}
    \includegraphics[height=0.45\textwidth,angle=-90]{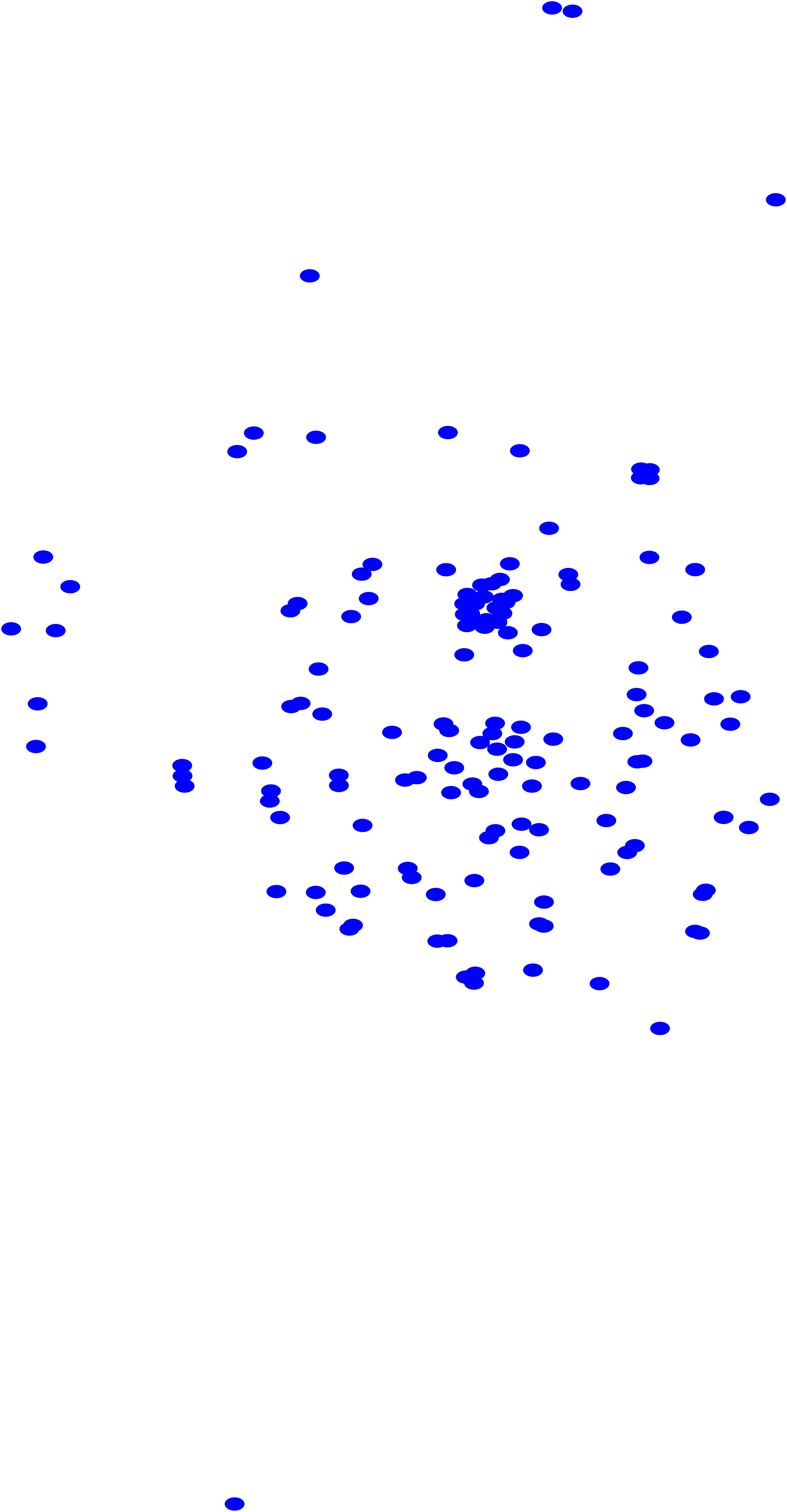}
    \vspace*{-12pt}
  \end{center}
  \caption{\it Connection graph of the 144 tests that are part of the
    \aspect{} code base.}
  \vspace*{-3mm}
  \label{fig:aspect-tests}
\end{wrapfigure}
Finally, there are cases where you copy-paste pieces from multiple existing
tests into a new one, for example because you want to test the combination of
features for which you have already written individual testcases. At this
point, the metaphor of inheritance graphs starts to break down. However, in
practice, most tests are still closely related to some others. A visualization
of this for the 144 tests that are part of \aspect{}, each represented by a
dot, is shown in Fig.~\ref{fig:aspect-tests} where we draw tests farther apart
if there are more textual differences between the input files. Many tests come
in groups of 2, 3 or 4 that only differ in a few lines of input -- simple
modifications such as those between testcases 3, 4 and 5 above. On the other
hand, other group are further apart and are derived from different
examplars.

The message from this section, though, is that in almost all cases writing a
new testcase is actually really easy: find a good starting point, copy it,
make minor modifications and additions, and verify the output. Testing
scientific codes really doesn't have to be hard!

\subsection{Dealing with many tests}
\label{sec:many-tests}

In practice, once you have accumulated a number of testcases, you will
encounter practical difficulties. First, when you try to find the right
starting point to clone and graft a new testcase, you will not recall if
\texttt{testcase-3.json} or \texttt{testcase-5.json} is the one
that's closest to the situation you have in mind (for example, if you want to
augment your model and you are looking again for a testcase with equal masses
and friction coefficients). 

In our experience, using more descriptive filenames for the input files has
helped, as is placing comments at the top of each input file describing what
the test does or, if appropriate, how it differs from a previous
testcase. Many of our testcases in \aspect{} have names like
\texttt{tests/melt-fraction-peridotite.prm} and comments at the top of the
form ``\textit{Like the melt-fraction.prm test, but using the peridotite
  material parameters.}''

Second, running and comparing many tests by hand using the commands above
becomes unwieldy. In such cases, it is useful to write little
\texttt{Makefile}s that do this automatically, or to use one of the tools
available as open source -- for example, the \texttt{CTest} program. It also
becomes more difficult to display tests that create differences when, for
example, you run the 7,000 tests of the \dealii{} testsuite. Again, there are
freely available systems such as \texttt{CDash} that present graphical
overviews of all succeeding and failing tests in a browser.

In practice, you're not likely to quickly get to a project size where learning
to use tools like \texttt{CTest} and \texttt{CDash} are important
considerations. In practice, the point where using such tools becomes
indispensible is when your code has several 10,000 lines of code or more --
several years worth of work. When you get to this point, your project will
have all sorts of other growing pains as well, such as having to deal with
external users, running mailing lists, portability, managing complexity and
other issues. Some of these are discussed in \cite{BH13}. The point simply is
that there is no need to worry too much about testing tools for programs of
this size: care about writing tests from the beginning and if the number of
tests becomes too unwieldy, there will be professional-grade tools you can
switch to that can manage this for you. In the meantime, a little shell script
such as the following will do just fine:
\begin{lstlisting}[frame=single,basicstyle=\footnotesize,numbers=left,language=bash]
# loop over all .json input files
for jsonfile in *.json ; do
  # execute our program on it and pipe the output into a file
  testname=`basename $jsonfile .json`
  python ../spring.py ${jsonfile} > ${testname}.output

  # then compare output; only ask 'diff' to report whether the
  # files are the same or not, and create output depending on this:
  # print an 'X' if there is a difference, a '.' if there isn't. this
  # ensures that test failures are easily visible
  if (diff ${testname}.reference ${testname}.output > /dev/null) ; then
    echo " .    ${jsonfile}" ;
  else
    echo " X    ${jsonfile}" ;
  fi
done
\end{lstlisting}
If you call this script, say, \texttt{tests/compare.sh}, then running it using
the command \texttt{cd tests/ ; bash -e compare.sh} should produce output like
this:
\begin{lstlisting}[frame=single,basicstyle=\footnotesize,numbers=left,language=bash]
 .    testcase-1.json
 .    testcase-2.json
 .    testcase-3.json
 .    testcase-4.json
 .    testcase-5.json
\end{lstlisting}
On the other hand, if your output looks like this
\begin{lstlisting}[frame=single,basicstyle=\footnotesize,numbers=left,language=bash]
 .    testcase-1.json
 .    testcase-2.json
 .    testcase-3.json
 X    testcase-4.json
 .    testcase-5.json
\end{lstlisting}
then you know that you just broke some functionality that
\texttt{testcase-4.json} checks. Given how easy it is to run this script, you
want to run this script often to find new bugs as early as possible.

\subsection{Dealing with overlapping tests}

In an ideal world, every testcase would only check a single feature (such
testcases are then often called \textit{unit tests}). This way, if something
goes wrong with a change to the source code, a single test will fail and it
will be easy to identify what went wrong.

In practice, this is often difficult to achieve. For example, if we change the
implementation of the function that provides $f(t,y)$ above and break it, it
is likely that \textit{all} testcases will start to fail. Similarly, if you
build testcases by cloning and grafting onto existing ones, whenever the old
one fails, the new one will likely do so as well.

From experience, it turns out that this is not often a real problem. The vast
majority of patches concern just small, isolated pieces of the software and in
those cases, only one or a few testcases will start failing. It is then easy
to identify the simplest one of these (assuming that they were given useful
names, see above in Section~\ref{sec:many-tests}), and that is the one we
will then use for debugging. If one breaks fundamental functionality somewhere
in the core of the software, a large number of tests will start to fail
but in those cases you will either be able to quickly find out what is
happening, or be able to debug the problem by choosing \textit{any} simple
test that was among those that broke. In summary, from a practical
perspective, we have found that it is not usually a problem to have testcases
that check overlapping parts of the software. Rather, we have come to the
conclusion that it is best to have as many tests as possible, regardless of overlap.

\subsection{Dealing with round-off}

Scientific computing deals, at a fundamental level, almost always with
floating point numbers. Unfortunately, floating point arithmetic suffers from
some rather annoying oddities that can make testing scientific codes more
challenging than code that doesn't use floating point numbers.

At the base of the problem is the fact that floating point numbers do not
enjoy the same mathematical properties as integers (or, in fact, as real
numbers). For example, if you take into account round-off, then $(a+b)+c$ is in
general \textit{not} the same as $a+(b+c)$ or $(a+c)+b$. You can try this out
with $a=1.0, b=-1.0, c=1e-20$: in double precision, you will get
$(a+b)+c=1e-20, a+(b+c)=(a+c)+b=0$. This matters because if you make changes
to the code base, upgrade the compiler, switch from one processor or machine
to another, or change compiler optimization level, floating point operations
will be executed in different order and the results will be different at the
level of roundoff. In other words, your tests will fail even though
nothing fundamentally changed.

Over the past 15 years, we have played with many ideas how to address this
problem. For example, we tried to just never output numbers to their full 16
digits of double precision. The idea is that if you only output 5 digits, then
roundoff surely can't play a role. The problem is that the numbers 1.234\,549
999\,999\,999\,9 and 1.234\,550\,000\,000\,000\,1 only differ by round-off,
but when rounded to 5 digits, you either get 1.234\,5 or 1.234\,6. One could
think that this is a contrived example, but it turns out that this happens
with rather frustrating regularity if you just output enough numbers -- as one
often does in scientific codes.

A second area that creates problems is that codes often try to compute zeros, but
because of roundoff what is actually produced are numbers like
4.863e-19. A human recognizes that this is a zero and, consequently, equal to
2.489e-19 if all of the numbers around it are on the order of one. But the
\texttt{diff} command will complain, and consequently signal a difference.

The solution to these problems that we have ultimately come to use
is a program called \texttt{numdiff}.%
\footnote{See \url{http://www.nongnu.org/numdiff/}.}
\texttt{numdiff} works essentially like
\texttt{diff} but it recognizes if the text it compares in two files is
actually a number. If two numbers differ by less than some relative or
absolute tolerance, then they are considered equal even if the characters by
which they are represented in an output file are different. For example, you
would replace the \texttt{diff} command used previously by the
following to ensure that numbers $a$ and $b$ compare as equal if $|a-b|\le
10^{-6}$ or $|a-b|\le 10^{-8} \min\{|a|,|b|\}$:
\begin{lstlisting}[frame=single,basicstyle=\footnotesize,numbers=left,language=csh]
  numdiff -a 1e-6 -r 1e-8 testcase-1.reference testcase-1.output
\end{lstlisting}
Using this trick, numbers that are written into the output file
\texttt{testcase-1.output} on one machine with slight differences to the
ones originally saved in \texttt{testcase-1.reference} will still be considered
the same, and the comparison will succeed.

\section{Summary}
\label{sec:conclusions}

When we talk to students who just start writing their own codes, then
they typically understand the importance of testing in some abstract
sense. But almost all of them have two fundamental misconceptions: First, that
their codes are either written well enough or will not be changed in ways
fundamental enough to suggest that these changes might break existing
functionality. And second, that doing adequate testing is just too hard and
cumbersome to be a justifiable expense of time.

The first of these two points is empirically false. Only bad programmers
expect that they will not break existing functionality, no matter how trivial
a change might be. All experienced programmers will know this to be wrong. The
reason is that over the lifetime of a scientific code -- say, over the course of
the 3 years a graduate student does research, and maybe even beyond this if
the code is handed on to the next student, or if you go on to be a postdoc --
a code starts out small but is constantly developed into something much larger
with much more functionality in all sorts of directions. You're not going to
create one essential version of the code that you can verify to be correct and
then just round off some edges; you're in fact going to replace the linear by
a nonlinear model, replace the direct solver by an iterative one, add terms to
the equation, and make other very fundamental changes that will require you to
edit all throughout your code. There is no other way than systematic testing
to ensure that the computation you ran today will still run tomorrow. And you
will have to run it again tomorrow: your adviser or boss may ask you to redo a
computation with some slight changes, a reviewer of a paper might ask to do
additional simulations, you may have forgotten the exact parameters for a data
set you want to use in your thesis and will have to redo it to verify,
etc. You may think that once you have created the data 
for some particular simulation, that the issue is done and will not have to
repeated -- i.e., that you might in fact just throw away the code -- but this
is not typically true.

The second of the points above -- that testing is hard and time consuming --
is also empirically false. Once you start to do it systematically, all you
have to do is morph an existing input file into a new one that tests the
aspect you have just implemented, and add it to your testsuite. It doesn't
take very long, you just have to do it right from the start. Once you do this,
you will find how much faster it suddenly becomes to debug codes. To us, one
of the most surprising facts of developing stringent testing procedures is not
just how much better our codes become, but in fact how much safer we feel
making changes in central places of our code: without tests, we always felt
insecure and wondered what a change might do to the code, whether it is
correct in all cases, and if we might be forgetting some cases. With tests,
we find ourselves a lot more at ease: just make a change, run the 140 or so
tests and see what happens -- if it all continues to work, then that may not
be a guarantee that a change is correct, but it sure makes you sleep better at
night. It really isn't all that hard to find peace of mind!

\paragraph*{Acknowledgments}
B.~Turcksin and W.~Bangerth were partially supported by the National Science
Foundation under award OCI-1148116 as part of the Software Infrastructure for
Sustained Innovation (SI2) program; and by the Computational
Infrastructure in Geodynamics initiative (CIG), through the National Science
Foundation under Award No.~EAR-0949446 and The University of California --
Davis.

T.~Heister was partially supported by the Computational
Infrastructure in Geodynamics initiative (CIG), through the National Science
Foundation under Award No.~EAR-0949446 and The University of California --
Davis.

\bibliographystyle{alpha}
\bibliography{paper}

\end{document}